\documentclass[conference]{IEEEtran}
\usepackage{amsmath}
\usepackage{amsthm}
\usepackage{amsfonts}
\usepackage{amssymb}
\usepackage{rotating}
\usepackage[bookmarks=false]{hyperref}
\hypersetup{
draft,
colorlinks=true,
linkcolor=blue,
filecolor=magenta, 
urlcolor=cyan,
}

\usepackage{tikz}
\usepackage{tabularx}

\usepackage{cite}

\ifCLASSINFOpdf
\else
\fi

\usepackage{stfloats}
\begin{document}
%
\title{Method to Annotate Arrhythmias by Deep Network}

\author{\IEEEauthorblockN{Weijia Lu\IEEEauthorrefmark{1}}
\IEEEauthorblockA{\IEEEauthorrefmark{1}Digital, General Electric\\
Shanghai, China\\
Email: AlfredWJLu@gmail.com}
\and
\IEEEauthorblockN{Jie Shuai\IEEEauthorrefmark{2} \hspace{0.5cm} Shuyan Gu\IEEEauthorrefmark{2}}
\IEEEauthorblockA{\IEEEauthorrefmark{2}Healthcare, General Electric\\
Shanghai, China}
\and
\IEEEauthorblockN{Joel Xue\IEEEauthorrefmark{3}}
\IEEEauthorblockA{\IEEEauthorrefmark{3}Healthcare, General Electric\\
Wauwatosa, WI, US}
}

%


\maketitle

\begin{abstract}
This study targets to automatically annotate on arrhythmia by deep network. The investigated types include sinus rhythm, asystole (Asys); supraventricular tachycardia (Tachy); ventricular flutter or fibrillation (VF/VFL); ventricular tachycardia (VT).\\
Methods: 13s limb lead ECG chunks from MIT malignant ventricular arrhythmia database (VFDB) and MIT normal sinus rhythm database were partitioned into subsets for 5-fold cross validation. These signals were resampled to 200Hz, filtered to remove baseline wandering, projected to 2D gray spectrum and then fed into a deep network with brand-new structure. In this network, a feature vector for a single time point was retrieved by residual layers, from which latent representation was extracted by variational autoencoder (VAE). These front portions were trained to meet a certain threshold in loss function, then fixed while training procedure switched to remaining bidirectional recurrent neural network (RNN), the very portions to predict an arrhythmia category. Attention windows were polynomial lumped on RNN outputs for learning from details to outlines. And over sampling was employed for imbalanced data. The trained model was wrapped into docker image for deployment in edge or cloud.\\
Conclusion: Promising sensitivities were achieved in four arrhythmias and good precision rates in two ventricular arrhythmias were also observed. Moreover, it was proven that latent representation by VAE, can significantly boost the speed of convergence and accuracy.\\

\end{abstract}


%
\IEEEpeerreviewmaketitle



\section{Introduction}\label{problem}
\indent Automatically annotate arrhythmia with deep learning network has been emerging along with the boost of diverse frameworks. Back to 2015, 1D convolutional neural networks (CNN) were proposed by Kiranyaz to classify ECG beats \cite{kiranyaz2016real}. In the same year, a stacked denoising autoencoders (DAE) combined with supervised classification was proposed by Rahhal for active learning the beat annotation \cite{al2016deep}. In past 2017, Ng and his group trained a CNN classifier as deep as 34 layers to predict 14 outputs \cite{rajpurkar2017cardiologist}. All these studies claimed a better performance compared with handcrafted feature, and even better than cardiologists annotation. \\
\indent This study targets on the same problem, in which a brand-new deep network is proposed. Data introduced in section II-A is partitioned into two parts for training and testing separately, preprocessed by typical filtering, transferred into 2D spectrum, and fed directly into this network. The network is trained by two steps, firstly feature vector in each time spot is summarized into latent representation by unsupervised learning, which acting as a warm initialization for further classifier training; secondly a classifier is retrieved by supervised learning from these latent representation. In the last part of this study, performance on test set is reported, and a comparative experiment is introduced to prove the pros of latent representation. 

\section{Method}\label{contents}
\subsection{Data}\label{data}
\indent ECG records are mainly from MIT Physionet database \cite{goldberger2000physiobank}, including malignant ventricular arrhythmia database (VFDB) \cite{greenwald1986development} and inbuilt normal sinus rhythm database. 

\begin{table}[!hp]
\centering
\caption{Rhythm of interest and their definition} 
\begin{tabular}{r | l}
\hline
Type & Definition\\
\hline
Normal sinus & normal heart rhythm \\
Asys & no rate at least 4s \\
Tachy & rate $>$ 140bpm for 17 beats \\
VF/VFL & F-wave \\
VT & rate in 100-250bpm, QRS span larger than 0.1s\\
\hline
\end{tabular}
\label{tab:rhythmDefinition}
\end{table}

The rhythm of interest includes asystole (Asys), supraventricular tachycardia (Tachy), ventricular flutter or fibrillation (VF/VFL) and ventricular tachycardia (VT); the corresponding definition can be found in table \ref{tab:rhythmDefinition}.

\subsection{Preprocessing}\label{preprocessing}
\indent A proper chunk of limb lead signal is firstly selected (e.g. 13s) and re-sampled (e.g 200Hz); the chunk span should be set based on the definition of arrhythmia in section II-A, and enclose segmentation around annotation label indicating rhythm change. Then a high passed FIR filter is employed to remove baseline wandering (e.g. -24dB at 0.05Hz); Finally 2D spectrum is computed based on welch's method (e.g. 1024 points of fast Fourier transform operated on 91\% overlapped moving window with span of 60 samples).


\subsection{Network to learn the category}\label{net}
\begin{figure*}[!b]
\centering
\includegraphics[width=0.85\textwidth]{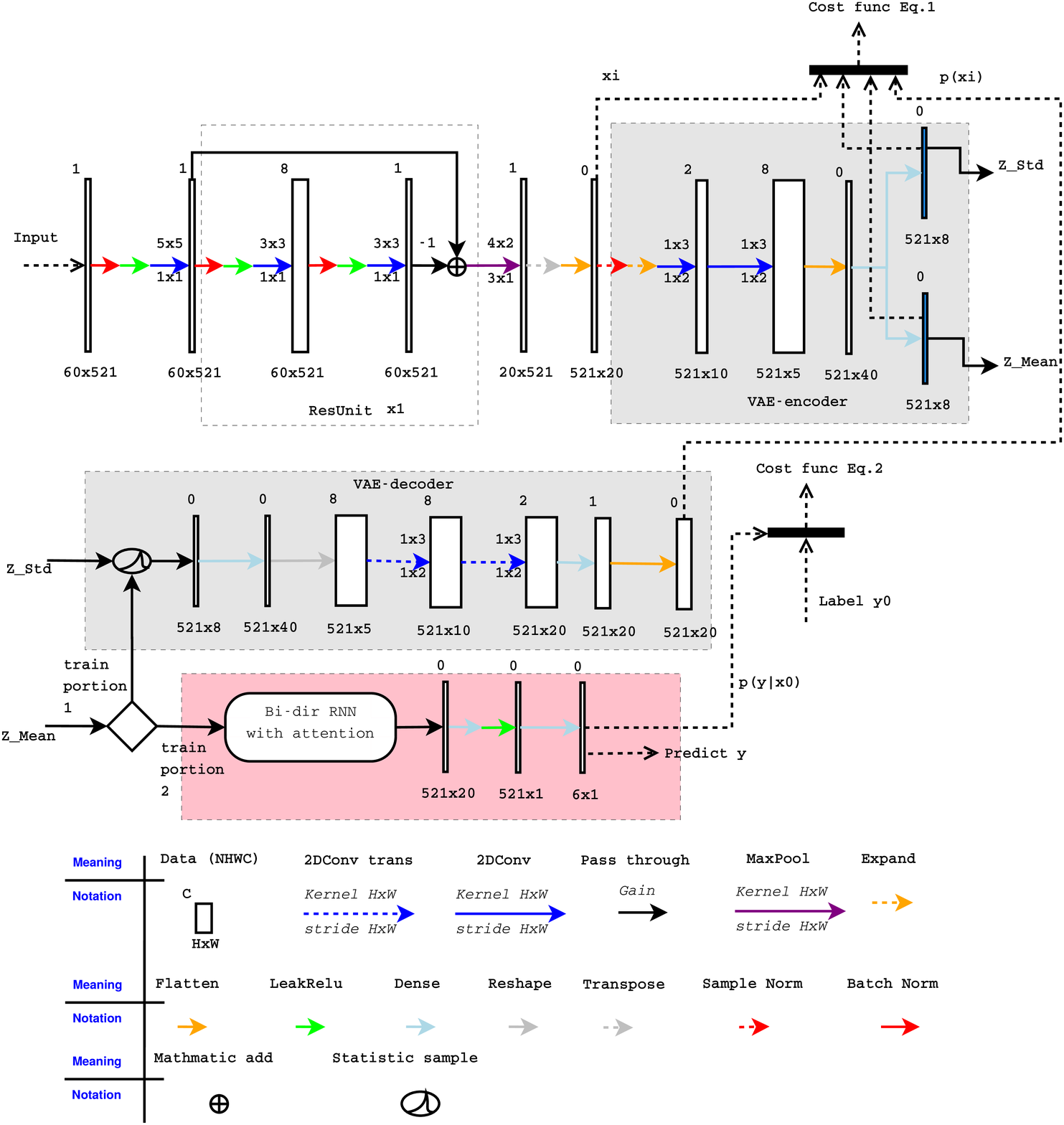}
\caption{Structure of proposed deep network. The highest dim of each data block represents size of batch (N), which is hidden in this diagram. The other three dimensions are height (H), width (W) and feature channels (C), the size of which can be read from footnote on each data box. For instance, first block shows a data batch with shape equal to $N (=140) \times 60 \times 521 \times 1$, each sample is a 2D spectrum, with 1 gray channel, 60 frequency bins and 512 sampling points.}
\label{netDiag}
\end{figure*}

\indent To annotate preprocessed signal, a network in Fig.\ref{netDiag} is proposed. This novel net structure, to our best knowledge, is never found in relative literature before. Therefore it requires an end-to-end training instead of transfer learning for most of image AI task. The whole network includes two parts, which need to be trained separately. The first one is a deep representation net (1425 floats), trained by cost function Eq.\ref{eq:cost2}. 
\begin{equation}
loss_{1}=E_{z\sim q_{\theta}(z|x_{i})}[|x_{i}-p_{\phi}(\widetilde{x_{i}})|^{2}|z] + KL(q_{\theta}(z|x_{i})||p(z|x))
\label{eq:cost2}
\end{equation}
This part gets a feature vector (20 floats) from frequency vector (60 frequency bins) by residual unit (ResUnit)\cite{he2016deep} and pooling, then tries to find a concise and robust representation of feature, drawn from Gaussian distribution in dense space (dimension = 8). The projection to dense space is retrieved by a variational autoencoder (VAE)\cite{kingma2013auto}, namely a pair of encoder-decoder highlighted in gray box in Fig.\ref{netDiag}.\\
\indent Back to mathematics in Eq.\ref{eq:cost2}, encoder is denoted by $q_{\theta}(z|x_{i})$ and decoder is represented as $p_{\phi}(\widetilde{x_{i}}|z)$, similar to the definition in Kingma et al's work\cite{kingma2013auto}. Therefore the first term in Eq.\ref{eq:cost2} is normally referred as reconstruction loss, for a latent representation $z$ is drawn and used to reconstruct original input $x_{i}$. The second term in Eq.\ref{eq:cost2} is Kullback-Leibler divergence when approximate the posterior $p(z|x)$ with a family of distributions $q_{\theta}$, and it acts as a regularizer penalty to maintain latent distribution into sufficiently diverse clusters\cite{kingma2013auto}. Therefore this term is normally referred as latent loss.\\
\indent The second portion, denoted by pink box in Fig.\ref{netDiag}, is a deep classification net (11790 floats), trained by cross entropy loss function Eq.\ref{eq:cost1}. 
\begin{equation}
loss_{2}=-E(\sum_{y=1}^{M}I_{y=y_{0}}log(p_{y|x_{0}}))
\label{eq:cost1}
\end{equation}
where $y$ and $y_{0}$ is predicted and true label respectively, M is total category of labels, $I(\bullet)$ is indicator function and $p_{y|x_{0}}$ is predicted distribution. This part is compounded by a multilayer bi-directional RNN structure along with multiple dense layers. The attention window, introduced by Bahdanau et al \cite{bahdanau2014neural}, is polynomial lumped on RNN cell outputs except last layer, namely the span of attention window is increasing from bottom to top layer. Therefore, the model is learning from details to overall picture. This portion reuses the output from first portion, and gives a prediction on each sample.\\
\indent In most cases, VAE serves as a traditional generative modeling. In our case, it's used to generate a robust latent representation and decrease noise in each sample, since as denoted in Kingma et al's study \cite{kingma2013auto}, VAE brings a constrain on the distribution of each latent cluster, and learns the distribution instead of a deterministic representation over training set. In later section a comparative test is presented to prove the effectiveness of VAE representation in this study.\\ 
\indent Fig.\ref{netDiag} shows a 'base' network configuration, including : kernel and stride size of filtering marked on both sides of operator arrow; number of residual units (=1); dimension of latent representation (=8); hidden unit size in RNN cell (=10); number of layers in bidrectional RNN network (=4) and smallest attention window size lump on RNN output (=3). Later on, this 'base' setup will be modified in the very same comparative test. 



\subsection{Method to train the network}\label{train}
\indent The original dataset after preprocessing, namely gray images indicating 2D spectrum of signal are randomly divided into several subsets (e.g. 5-fold) in each arrhythmia type. Network are tested on a certain subset after aggregating those samples from different arrhythmia types. On the other hand the remaining data are organized into mini batch and fed directly into model for training purpose, after over sampling in under-represented arrhythmia categories. Such design of data preparation are specific for imbalanced set in this study, namely the big difference of total samples falling in each arrhythmia categories. The mini batch size is equal to 140 in this study.\\
\indent As mentioned in section \ref{net}, the whole net is trained separately. The training of first part is fulfilled after a certain threshold of cost is met, then the first part is fixed and model start to be trained on second part. In VAE, the cost function is combined by two mathematics terms as in Eq.\ref{eq:cost2}. From these two parts, a fraction $\eta$ can be calculated to represent the portion of latent loss in total loss in each iteration. This fraction is coupled into the model training procedure in this study to dominate sampling in latent Gaussian distribution. In the very beginning, the reconstruction loss takes a large portion in total loss, therefore the fraction is small and latent variable $z$ is drawn from normal distribution $N(\mathbf{\scriptstyle Z\_Mean},\eta \cdot \mathbf{\scriptstyle Z\_Std})$ during sampling instead of $N(\mathbf{\scriptstyle Z\_Mean},\mathbf{\scriptstyle Z\_Std})$. By this manner, model convergences faster to a potential cluster center in several epochs and total loss decreases rapidly. Later on, latent loss gradually takes a dominant portion, and regularizer effect starts working.

\subsection{Method to deploy this model}\label{deploy}
\indent The trained model is encapsulated into docker image, and deployed to either edge computer server or virtual private cloud (VPC). During the initialization of docker container in edge server, local working directory can be mounted, and data can be fed into model by local file I/O. As for the case in VPC, web RESTful API can be provided by flask or likely python component as major data I/O.

\section{Experiment and result}\label{result}

\begin{figure*}[!b]
\centering
\begin{tabular}{cc}
\includegraphics[width=0.8\columnwidth]{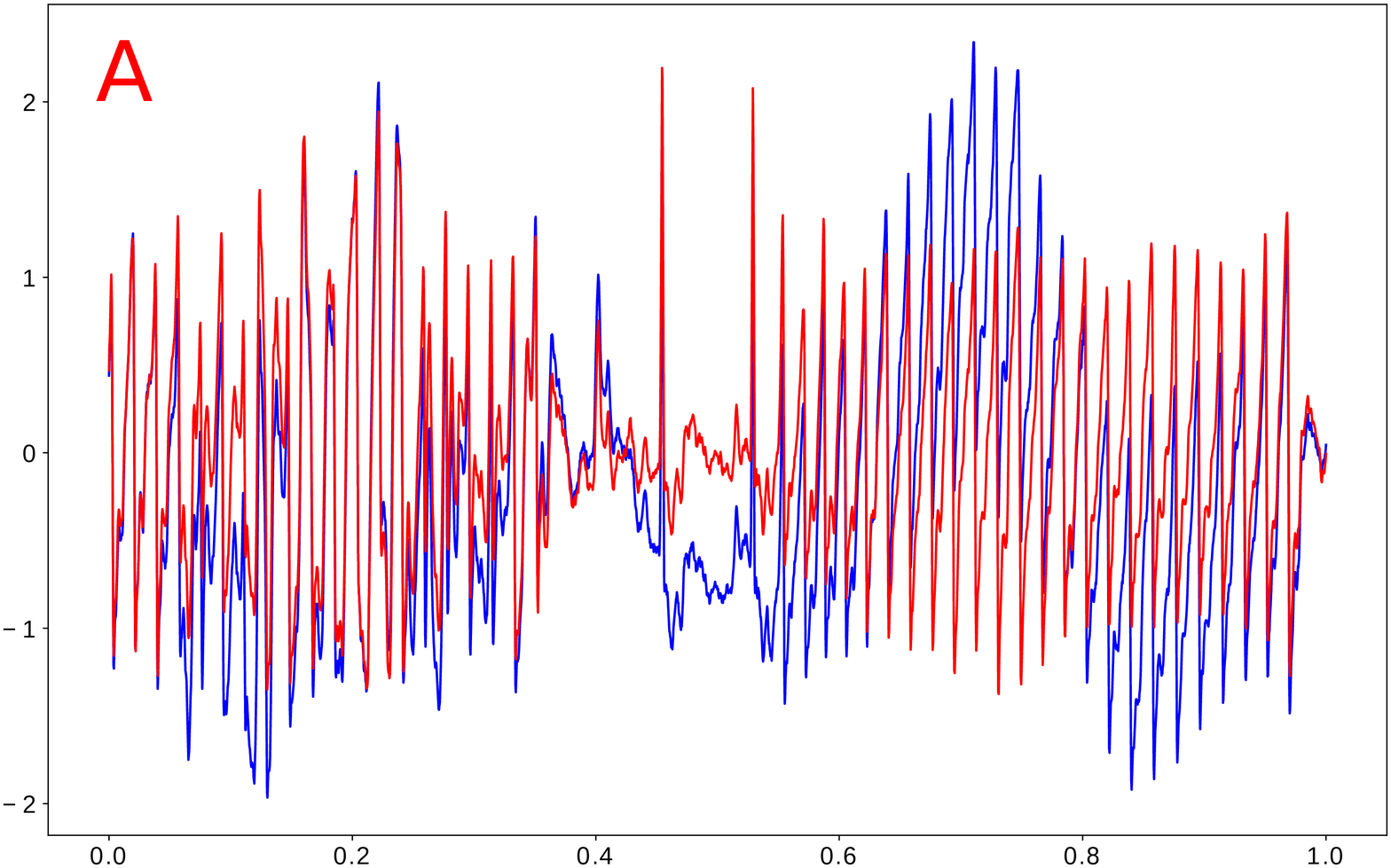} &
\includegraphics[width=0.8\columnwidth]{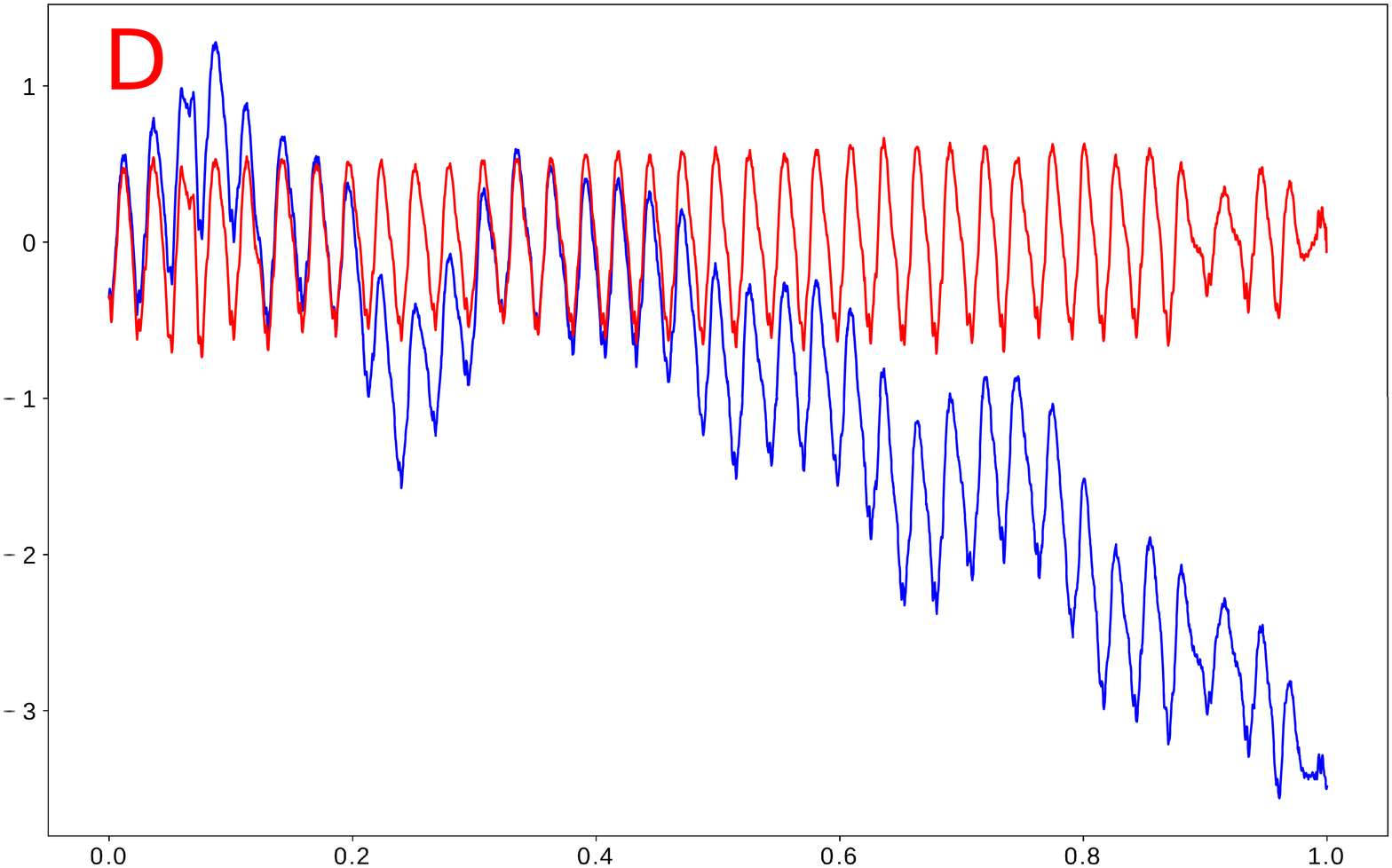} \\
\includegraphics[width=0.9\columnwidth]{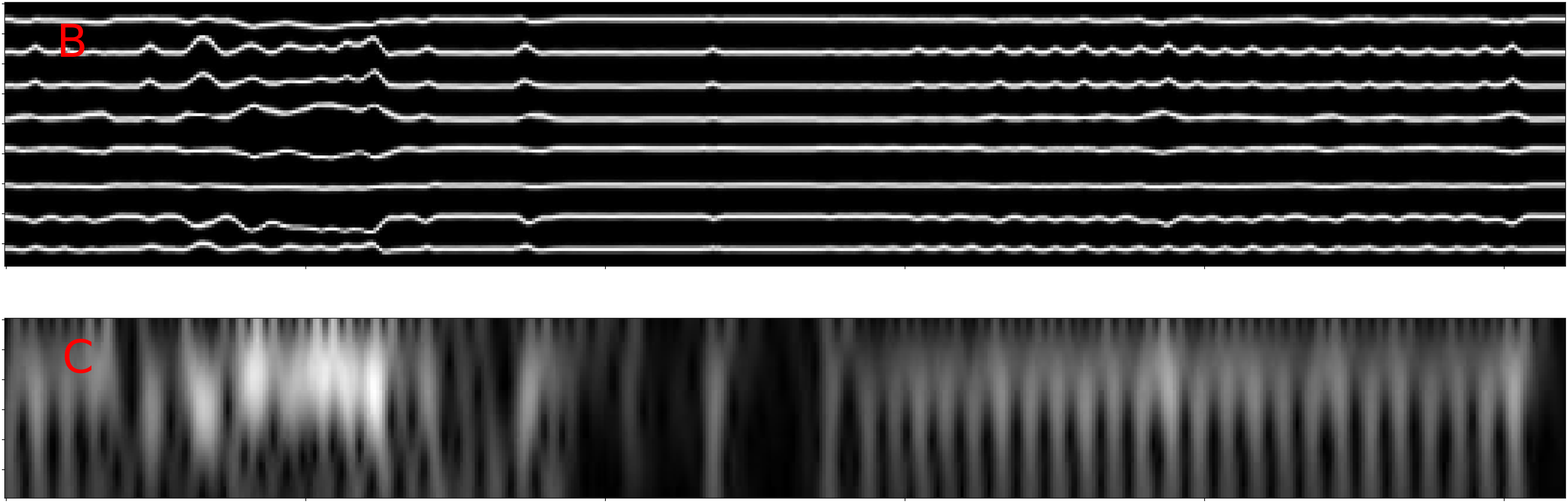} &
\includegraphics[width=0.9\columnwidth]{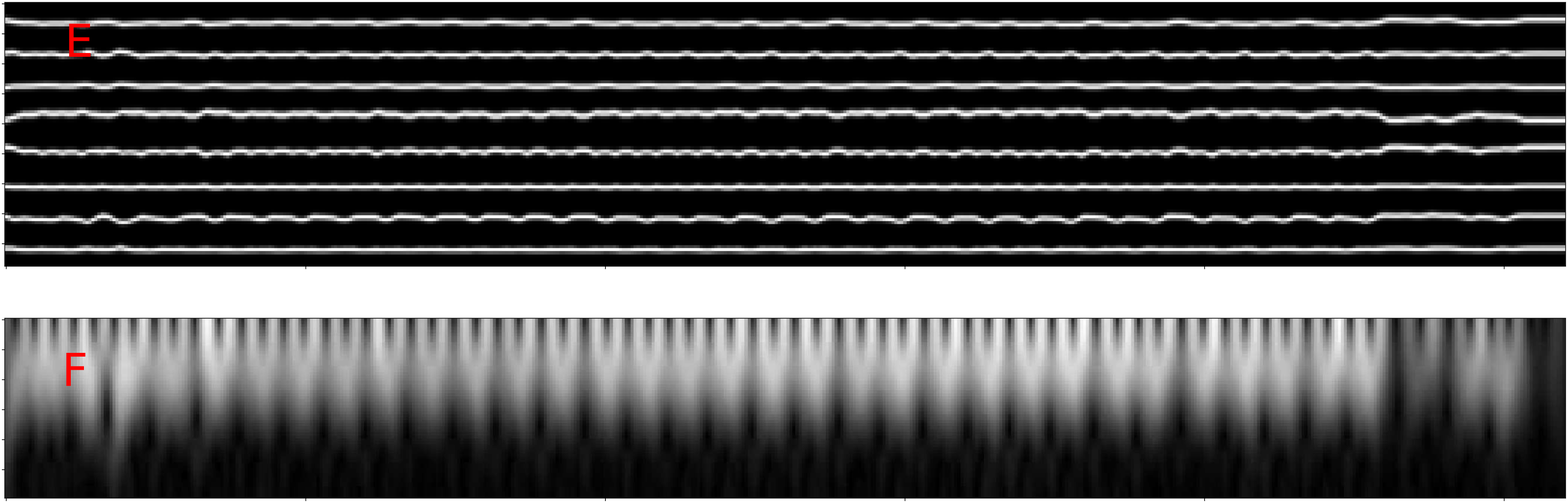}
\end{tabular}
\caption{Original and preprocessed ECG waveform along with its 2D spectrum and latent representation. (a-c) A VT sample $\#421.a865$, its 2D spectrum and its latent representation correspondingly; (d-f) a supraventricular tachycardia sample $\#426.a080$, its 2D spectrum and latent representation. The dimension of latent space equal to 8 in this setup. In the latent representation, each line represents an coordinate of latent space. Its amplitude gives the mean, and its width gives the standard deviation of a Gaussian distribution. Thus the whole picture gives the propagation of latent representation on time.}
\label{lattentRep}
\end{figure*}

\begin{table}[!hpb]
\centering
\caption{Confusion matrix on \#0 VFDB test set in 5 fold CV design (accuracy = 90\%)} 
\begin{tabular}{r | c c c c c}
Type & Asys & Tachy & VF/VFL & VT\\
\hline
\# PP/Tot & 48/52 & 19/20 & 398/444 & 335/372 \\
Sensitivity & 0.92 & 0.95 & 0.89 & 0.90\\
Precision & 0.74 & 0.70 & 0.96 & 0.90\\
\end{tabular}\\ 
\label{tab:ConfusionMatrixOnVFDBTestSet}
\end{table}

\begin{table}[!hpb]
\centering
\caption{Confusion matrix on \#1 VFDB test set in 5 fold CV design (accuracy = 90\%)} 
\begin{tabular}{r | c c c c c}
Type & Asys & Tachy & VF/VFL & VT\\
\hline
\# PP/Tot & 46/52 & 16/20 & 399/444 & 345/372 \\
Sensitivity & 0.88 & 0.80 & 0.89 & 0.93\\
Precision & 0.82 & 0.72 & 0.94 & 0.89\\
\end{tabular}\\ 
\label{tab:ConfusionMatrixOnVFDBTestSet2}
\end{table}
\indent The model is trained and tested in 5-fold cross validation. Fig.\ref{lattentRep} shows the intermediate plotting on a VT sample and a Tachy sample, including: waveform before and after prefiltering, relative 2D spectrum and relative latent representation. Table \ref{tab:ConfusionMatrixOnVFDBTestSet} and \ref{tab:ConfusionMatrixOnVFDBTestSet2} give the performance rates on test sets using a so-called 'base' setup in section \ref{net}. From these tables, it's proven that proposed network has promising sensitivity in detection these arrhythmias, and good precision in two ventricular arrhythmias.\\

\begin{figure*}[!b]
\centering
\includegraphics[width=0.8\textwidth]{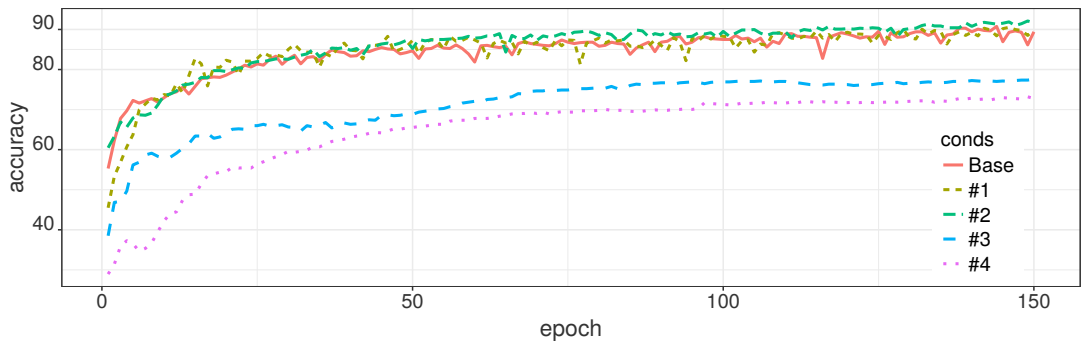}
\caption{Accuracy on test set under different setup during training. Comparing with setup termed as base, \#1 increases hidden units size in each RNN cell from 10 to 15; \#2 increases smallest attention window from 3 to 5, whistle decreases number of directional layers from 4 to 3; \#3 and \#4 has the same configuration comparing with \#1 and \#2 correspondingly, but use a dense projection instead of variational auto-encoder to get the latent representations. Therefore in \#3 and \#4, network is trained as a whole set.}
\label{compTest}
\end{figure*}

\indent Moreover, a comparative experiment is conducted, and accuracy of model on test set is recorded for comparative plotting. During this experiment, different setup on network parameter, e.g. the number of hidden units in RNN cell or span of attention window are tested; in addition to these superparameters, two methods to get latent representation are compared too : using VAE to learn a statistic distribution or retrieving a deterministic latent variable from a simple dense projection. The results can be found in Fig.\ref{compTest}. From this figure, it's proven that to learn a latent distribution by VAE can significantly boost the speed of convergence and accuracy. Meanwhile, other configuration in setup has no significant difference on the model performance. \\

\section{Conclusion}
\indent In this study, a brand-new deep network structure is proposed to annotate arrhythmia in ECG signal. This network includes two parts, one for extracting latent representation from feature vector in each time spot, and the other for predicting arrhythmia categories of interest. This network achieve around 90\% sensitivity in asystole, ventricular flutter/fibrillation or ventricular tachycardia, and over 80\% sensitivity in all 4 arrhythmias including supraventricular tachycardia. Moreover the test results show good precision rates in two ventricular arrhythmias. This network uses VAE to extract latent representation, which works as a warm initialization for classifier. This method is proven to significantly improve the speed of convergence and accuracy.


%
%



\bibliographystyle{IEEEtran}
%

\end{document}